\documentclass[aps,pra,nobalancelastpage,longbibliography,showpacs,twocolumn,superscriptaddress,floatfix]{revtex4-1}

\usepackage{amsmath,amssymb,graphicx, braket, float}
\usepackage{blindtext}

\counterwithout{figure}{section}
\usepackage[export]{adjustbox}
\graphicspath{{figure/}}

\usepackage[dvipsnames]{xcolor}
\usepackage[breaklinks=true,colorlinks=true,linkcolor=red,urlcolor=blue,citecolor=blue]{hyperref}

\usepackage{graphicx, xcolor}
\usepackage[margin=.8in,text={10in,10.5in},centering]{geometry}
\usepackage{subcaption}
\usepackage[T1]{fontenc}

\newcommand{\sx}[0]{\hat{\sigma}_\mathrm{x}}

\newcommand{\sz}[0]{\hat{\sigma}_\mathrm{z}}

\usepackage{orcidlink}

\usepackage[normalem]{ulem}
\newcommand{\away}[1]{\textcolor{red}{\ifmmode\text{\sout{\ensuremath{#1}}}\else\sout{#1}\fi}}

\begin{document}
	\title{Nonequilibrium quantum thermometry with noncommutative system-bath couplings}

	\author{Youssef Aiache\,\orcidlink{0009-0001-9294-138X}} \email{youssefaiache0@gmail.com}
	\address{Laboratory of R\&D in Engineering Sciences, Faculty of Sciences and Techniques Al-Hoceima, Abdelmalek Essaadi University, BP 34. Ajdir 32003, Tetouan, Morocco}
	
	\author{Abderrahim El Allati\,\orcidlink{0000-0002-2465-8515}}
	\address{Laboratory of R\&D in Engineering Sciences, Faculty of Sciences and Techniques Al-Hoceima, Abdelmalek Essaadi University, BP 34. Ajdir 32003, Tetouan, Morocco}
	
	\author{\.{I}lkay Demir}
	\affiliation{Department of Nanotechnology Engineering, Sivas Cumhuriyet University, 58140 Sivas, T\"{u}rkiye}
	\affiliation{Sivas Cumhuriyet University Nanophotonics Application and Research Center-CUNAM, 58140 Sivas, T\"{u}rkiye
	}  
	
	\author{Khadija El Anouz\,\orcidlink{0000-0002-6437-826X}}
	\address{Laboratory of R\&D in Engineering Sciences, Faculty of Sciences and Techniques Al-Hoceima, Abdelmalek Essaadi University, BP 34. Ajdir 32003, Tetouan, Morocco}

	%%------------------------
	\begin{abstract}
	%Low temperature estimation is 
	%Accurate temperature estimation in the low-temperature regime remains a central challenge in quantum thermodynamics. In this work, we explore nonequilibrium quantum thermometry using a single-qubit probe coupled to a bosonic thermal bath through noncommuting interaction operators. This framework unifies pure dephasing and dissipative processes within a spin–boson model and reveals how their interference profoundly alters thermal sensitivity. By tuning the coupling structure, we demonstrate that the probe’s temperature sensitivity exhibits a quadratic scaling, in the limit $ T \to 0 $, even under weak coupling. Furthermore, we show that coherence-based measurements, though suboptimal, become the most informative at early times. These findings highlight the role of noncommutative couplings as a resource for enhancing quantum thermometric performance and provide a physically grounded pathway toward experimentally realizable, nonequilibrium temperature probes.
	Accurate temperature estimation in the quantum and cryogenic regimes remains a fundamental challenge. Here, we investigate nonequilibrium quantum thermometry using a single-qubit probe coupled to a bosonic bath through noncommuting interaction operators, which unify pure dephasing and dissipative dynamics within a spin–boson model. We show that the interference between these two coupling channels induces strong non-Markovian feedback between populations and coherences, leading to coherence trapping and enhanced thermal sensitivity. Remarkably, by tuning the coupling structure, the probe’s temperature sensitivity exhibits a quadratic low-temperature scaling, even under weak coupling. Moreover, while coherence-based measurements are formally suboptimal, they become the most informative in the early nonequilibrium regime, where memory effects dominate. Our findings identify noncommutative system–bath couplings as a practical and tunable resource for achieving high-precision quantum thermometry in realistic open-system architectures.
	
	\end{abstract}
	
	\maketitle

	\section{Introduction}
	\label{Intro}
	High-precision temperature estimation at the nanoscale holds major importance, ranging applications from fundamental natural sciences to the rapidly developing field of quantum technologies~\cite{brites2012thermometry,binder2018thermodynamics,mehboudi2019thermometry}. Indeed, recent advances in microscopic physics shed light on quantum thermodynamics in which the precise control and measurement of thermodynamic quantities at the quantum level necessitates quantum-based methods, since classical approaches frequently prove ineffective~\cite{brandao2015second,deffner2019quantum,pekola2015towards,alicki2019introduction,whitney2019quantum,horodecki2013fundamental,khoudiri2025non,ait2025quantum}. In this inspiration, the so-called quantum thermometry aims to accurately determine the temperature of quantum systems. Indeed, it represents a general strategy for probing the temperature of a quantum reservoir via placing a quantum thermometer in thermal contact with it. Hence, according to their interaction, one can encode thermal information into the state of the thermometer. Then, the temperature can be inferred either from the equilibrium thermal state~\cite{latune2020collective,de2016local,ullah2024mixing} or from the nonequilibrium dynamical state~\cite{mok2021optimal,de2017estimating,seah2019collisional,mancino2020nonequilibrium,aiache2024non} of the thermometer by measuring suitable observables.\par

	Quantum thermometry can surpass classical approaches by exploiting quantum coherence~\cite{stace2010quantum,jevtic2015single,frazao2024coherence}, strong coupling~\cite{brenes2023multispin,mihailescu2023thermometry,rodriguez2024strongly}, quantum correlations~\cite{gebbia2020two,planella2022bath}, and periodic driving~\cite{mukherjee2019enhanced,glatthard2022bending,xu2023non}. However, a major challenge in most schemes reflect the sensing error typically diverges as the temperature decreases. Indeed, strategies such as strong coupling, periodic driving, and finite measurement resolution can decrease this divergence~\cite{potts2019fundamental}, where reliable performance in the ultra-low temperature regime remains elusive. In the thermalized case, universal results link thermometric precision to the heat capacity via the Gibbs ensemble. But, for very low temperatures, quantum probes often fail to thermalize and instead reach model-dependent nonthermal steady states. For equilibrium probes, the estimation error diverges exponentially as $T \to 0$, whereas nonequilibrium probes can reduce this to a polynomial divergence, though the error can still be substantial.\par
	
	In this work, we investigate the process of temperature estimation in a bosonic thermal bath using a single-qubit thermometer evolving under non-Markovian dynamics. Our main goal is to elucidate how the structure of the system–bath interaction, implemented through noncommuting coupling operators, affects thermometric sensitivity. This framework unifies the limiting cases of pure dephasing and purely dissipative interactions and allows one to explore the full range of intermediate couplings, where interference between these channels becomes operational. We show that such noncommutative couplings generate a dynamical feedback between populations and coherences of the probe, leading to long-lived coherence trapping and an enhancement of temperature sensitivity. In particular, we find that in the low-temperature limit, $ T \to 0 $, the sensitivity exhibits a quadratic scaling, even under weak coupling, whereas conventional Born–Markov treatments predict an exponential suppression. Finally, we demonstrate that coherence-based measurements, while formally suboptimal, become the most informative at early nonequilibrium regime, regimes that cannot be captured by Markovian master equations~\cite{de2017dynamics}, where non-Markovian memory effects are most pronounced. This highlights noncommutative couplings as a resource for enhancing the precision and robustness of quantum thermometry in experimentally realistic open-system settings.\par
	The paper is organized as follows. Section~\ref{NQT} describes the model, its dynamical properties, and provides a brief overview of the temperature estimation formalism. Section~\ref{R_and_D} presents and discusses the results. Section~\ref{Experimental_Outlook} outlines the experimental outlook of our proposed model. Finally, Section~\ref{Concl} summarizes the main conclusions of the paper.
	\section{Non-equilibrium quantum thermometry}
	\label{NQT}
	%\subsection{INFORMATION METRICS}% Thermometeric performance
	%\label{IM}
	
	%\subsection{OUR SCHEME}
	%\label{scheme}
	\noindent\textit{The model---} We consider a single qubit, i.e., a thermometer (we will use the terms interchangeably) of the following Hamiltonian:  
	\begin{equation}
		\hat{H}_S = \dfrac{\hbar~\varepsilon}{2}\sigma_z,
	\end{equation}
	where $ \varepsilon $ is the spin splitting. Whereas, the sample to be probed is a bath of a non-interacting bosonic system that rests at thermal equilibrium at a temperature $ T $. Then, the Hamiltonian of the bath reads as: 
	\begin{equation}
		\hat{H}_{B}=\sum_{k}\hbar\omega_k \hat{b}_k^{\dagger} \hat{b}_k,
	\end{equation}
	where $\omega_k$ is the frequency of the reservoir modes, $ \hat{b}_k^{\dagger}(\hat{b}_k) $ defines the bosonic creation (annihilation) operator for mode $ k $. However, the thermometer is coupled to the sample using the following linear interaction:  
	\begin{equation}
		\hat{H}_{int}=\hat{\sigma}_{\alpha}\otimes\sum_{k}\hbar~g_k( \hat{b}_k^{\dagger}+ \hat{b}_k)=\hbar\hat{\sigma}_{\alpha}\otimes B,
	\end{equation}
	where $ g_k $ denotes the strength coupling of each mode with the thermometer. Moreover, it is common practice to incorporate the frequency dependence of the interaction strengths into the spectral density, which is defined as 
	$J_{\omega} \equiv \sum_k g_k^2\, \delta(\omega - \omega_k)$~\cite{breuer2002theory,vacchini2024open}. However, in this work, we adopt an Ohmic spectral density of the following form: 
	\begin{equation}
			J_{\omega} = \eta\, \omega\, e^{\omega/\omega_c},
	\end{equation}
	where $\eta$ defines a dimensionless coupling constant that carries the order of magnitude of the couplings $ g_k $. While, $\omega_c$ is the cutoff frequency.\par 
	We consider the role of noncommutative coupling~\cite{lu2013sub,duan2020unusual,zhang2021non} in the interaction between the thermometer and the sample by introducing a parameter $ \alpha $ that defines the structure of the coupling operator. Specifically, we define the coupling operator as: 
	\begin{equation}
		\label{int-oper}
	\hat{\sigma}_{\alpha} = (1 - \alpha)\, \sz + \alpha\, \sx,
	\end{equation}
	which allows interpolation between purely diagonal and purely off-diagonal interactions in the energy basis of the thermometer. In fact, for $ \alpha = 0 $, the coupling reduces to a purely dephasing interaction mediated by $ \hat{\sigma}_z $, which preserves the populations of the thermometer’s energy levels while affecting the coherences. Conversely, for $ \alpha = 1 $, the interaction is governed by $ \hat{\sigma}_x $, corresponding to a purely dissipative or energy-exchange coupling. This parametrization enables the exploration of intermediate regimes, where both diagonal and off-diagonal coupling effects coexist. This offers a flexible framework to investigate how the nature of the system–probe interaction influences the performance of quantum thermometry.\par
	Since we are interested only in the dynamics of the thermometer, we employ an exact non-Markovian master equation derived using Zwanzig’s projection-operator method~\cite{breuer2002theory,vacchini2024open,smirne2010nakajima} to trace out the sample degrees of freedom from the Liouville equation of motion, retaining terms up to second order in the thermometer–sample coupling.
	\begin{eqnarray}\label{NME}
		\dot{\rho}_{S}&=&i/\hbar~[ \rho_{S}, H_S]\\ &+&\intop_{0}^{t}d\tau\{\Phi_T(t-\tau)[\sigma_{\alpha}(t,\tau)~\rho_{S},\sigma_{\alpha}]+h.c\}\nonumber,
	\end{eqnarray}
	where $\Phi_T(t)=\langle e^{i H_B t} B e^{-i H_B t} B \rangle_{\rho_{B}} $
	is the bath correlation functions, with $ \langle \dots \rangle_{\rho_{B}}=\mathrm{Tr}\{\dots~\rho_{B}\} $. $\rho_B$ is the thermal equilibrium state of the reservoir at temperature $T$. Indeed, it takes the following form:
	\begin{equation}
	\rho_B = \prod_k (1 - e^{\beta \omega_k}) e^{-\beta \omega_k b_k^\dagger b_k},
	\end{equation}
	where $\beta = \hbar / k_B T$ is the inverse temperature ($\hbar=k_B=1$). The time-dependent interaction operator is expressed as 
	$ \hat{\sigma}_{\alpha}(t,\tau) = \alpha ~ \hat{\sigma} _-~ e^{i \varepsilon (t-\tau )}+\alpha ~ \hat{\sigma} _+~ e^{-i \varepsilon (t-\tau )}+(1-\alpha ) ~\hat{\sigma} _z $. Additionally, we assume that the initial state of the thermometer-sample system as $ \rho_{tot}(0) = \rho_{S}(0) \otimes \rho_{B} $, where the initial state of the thermometer is chosen as $ \rho_{S}(0)=\ket{+}\bra{+} $, such that $ \ket{+}=\dfrac{1}{\sqrt{2}}(\ket{0}+\ket{1}) $.\par
	It is worth emphasizing that the master equation~(\ref{NME}) is exact, where no Markov and even no secular approximations have been applied. Hence, the density matrix of the thermometer by means of a single qubit can be conveniently rewritten in the Bloch representation~\cite{nielsen2010quantum} as: $
	\rho_S = \frac{1}{2} \left( \mathbb{I}_2 + \mathbf{\Delta} \cdot \hat{\boldsymbol{\sigma}} \right) $, where $\mathbb{I}_2$ is the $2 \times 2$ identity matrix. While, $\mathbf{\Delta} = (\Delta_x, \Delta_y, \Delta_z)$ denotes the real Bloch vector characterizing the state of the qubit and $\hat{\boldsymbol{\sigma}} = (\hat{\sigma}_x, \hat{\sigma}_y, \hat{\sigma}_z)$ denotes the vector of Pauli matrices.
	Therefore, one can obtain the generalized Bloch equations for the components of the thermometer’s state as: 
	\begin{widetext}
		\begin{eqnarray}\label{Bloch_Eq}
			\dot{\Delta}_x &=& -\varepsilon\, \Delta_y - 4 \alpha (\alpha - 1) G_t - 4 \alpha (\alpha - 1)\, \Delta_z\, \mathcal{K}_t - 4 (\alpha - 1)^2\, \Delta_x\, R_t, \label{eq:dSxp} \\
			\dot{\Delta}_y &=& \Delta_x \left(\varepsilon + 4 \alpha^2 X_t \right) + 4 (\alpha - 1) \alpha (F_t - L_t) - 4 \Delta_y \left(\alpha^2 \mathcal{K}_t + (\alpha - 1)^2 R_t \right) - 4 (\alpha - 1) \alpha\, \Delta_z\, X_t, \label{eq:dSyp}\\
			\dot{\Delta}_z &=& -4 \alpha^2 G_t - 4 \alpha^2 \Delta_z\, \mathcal{K}_t - 4 (\alpha - 1) \alpha\, \Delta_x\, R_t, \label{eq:dSzp}
		\end{eqnarray}
	\end{widetext}
	where the time-dependent coefficients in the above equation are expressed as: 
	\begin{widetext}
		\begin{eqnarray}\label{Coeff}
			R_t&=&\int_0^{\infty } \frac{J_{\omega} \sin (t \omega ) \coth \left(\frac{\omega }{2 T}\right)}{\omega } \, d\omega,~~~~\mathcal{K}_t=\int_0^{\infty } \frac{J_{\omega} \coth \left(\frac{\omega }{2 T}\right) (\varepsilon \sin (\varepsilon t) \cos (t \omega )-\omega  \cos (\varepsilon t) \sin (t \omega ))}{\varepsilon^2-\omega ^2} \, d\omega,\nonumber\\
			L_t&=&\int_0^{\infty } \frac{J_{\omega} (1-\cos (t \omega ))}{\omega } \, d\omega,~~~X_t=\int_0^{\infty } \frac{J_{\omega} \coth \left(\frac{\omega }{2 T}\right) (-\omega  \sin (\varepsilon t) \sin (t \omega )+\varepsilon (-\cos (\varepsilon t)) \cos (t \omega )+\varepsilon)}{\varepsilon^2-\omega ^2} \, d\omega,\nonumber\\
			F_t&=&\int_0^{\infty } \frac{J_{\omega} (\varepsilon \sin (\varepsilon t) \sin (t \omega )+\omega  \cos (\varepsilon t) \cos (t \omega )-\omega )}{\varepsilon^2-\omega ^2} \, d\omega,~~~G_t=\int_0^{\infty } \frac{J_{\omega} (\omega  \sin (\varepsilon t) \cos (t \omega )-\varepsilon \cos (\varepsilon t) \sin (t \omega ))}{\varepsilon^2-\omega ^2} \, d\omega.\nonumber
		\end{eqnarray}
	\end{widetext} 
	The structure of the generalized Bloch equations Eqs.~(\ref{eq:dSxp}--\ref{eq:dSzp})
	provides a transparent physical picture of the thermometer’s dynamics under noncommutative couplings. The time-dependent coefficients $R_t, \mathcal{K}_t, X_t, F_t, L_t$, and $G_t$ represent dynamical memory kernels induced by the structured bath, capturing how past system–bath correlations influence the instantaneous evolution of the probe. Crucially, the terms proportional to $\alpha(1-\alpha)$ in Eqs.~(\ref{eq:dSxp}--\ref{eq:dSzp})
	are direct manifestations of quantum interference between the dephasing $(\sz)$ and dissipative $(\sx)$ interaction pathways. For intermediate values of $\alpha$, these noncommutative contributions encode a cross-coupling between the population $(\Delta_z)$ and coherence $(\Delta_x)$ dynamics. Meaning the bath simultaneously monitors the qubit’s energy (a quantum non-demolition process linked to dephasing) while inducing transitions between its levels (a dissipative process). The concurrent action of these two channels establishes a dynamical feedback, where coherences affect population flow and vice versa.
	%, a mechanism absent in the case of $\alpha=0$ and $\alpha=1$. 
	%This interference not only underlies the enhanced thermometric sensitivity observed at intermediate $\alpha$, but can also lead to partial coherence trapping or long-lived quantum coherence. The latter arises from the non-Markovian backflow of information sustained by the structured environment, allowing the probe to retain quantum correlations that would otherwise decay irreversibly in purely dephasing or dissipative regimes.

	\noindent\textit{Information backflow---} 
As it has already been mentioned, we aim to employ a non-Markovian master equation to investigate the role of the coupling mixing parameter $ \alpha $ on the thermometric performance. Indeed, we shall study the influence of $ \alpha $ on the emergence of non-Markovian dynamics in this model. In this regard, note that various definitions of non-Markovianity have been proposed in the literature~\cite{wolf2008assessing,breuer2009measure,smirne2022holevo,megier2021entropic,aiache2025non}. In this work, we adopt a coherence-based approach, where non-Markovianity is witnessed through re-coherence~\cite{chanda2016delineating,he2017non,wu2020detecting}, that is, the temporary revival of quantum coherence during the system's evolution. This revival indicates a reversal in the decoherence process, which gives rise to back-flow of information from the environment to the open system. Accordingly, the presence of re-coherence serves as a qualitative signature of non-Markovian behavior in the dynamics~\cite{breuer2009measure}.
	\begin{eqnarray}\label{NM}
		\mathcal{N}_C = \int_{\dot{C}(t) > 0} \dot{C}(t)\,dt, 
	\end{eqnarray}
where $ C(t) $ reflects the quantum coherence~\cite{streltsov2017colloquium,aiache2024dynamics} of the system, quantified by means of the so-called $ l_1 $-norm of coherence. Note that in Bloch representation, the $ l_1 $-norm of coherence can be expressed as $ C(t) = \sqrt{\Delta_x^2 + \Delta_y^2} $, where $ \Delta_x $ and $ \Delta_y $ are the transverse components of the Bloch vector obtained from Eq. (\ref{Bloch_Eq}). In general, if the coherence at $t=0$ satisfies $ C(0) = 1 $, then the state corresponds to a pure state on the equator of the Bloch sphere. However, the re-coherence is identified for distinct intervals when $ \dot{C}(t) > 0 $~\cite{he2017non}, where one can integrate successfully these revivals over time to obtain the total non-Markovianity witness, namely $ \mathcal{N} $.\par 
\begin{figure}[H]
	\begin{center}
		\includegraphics[scale=.42]{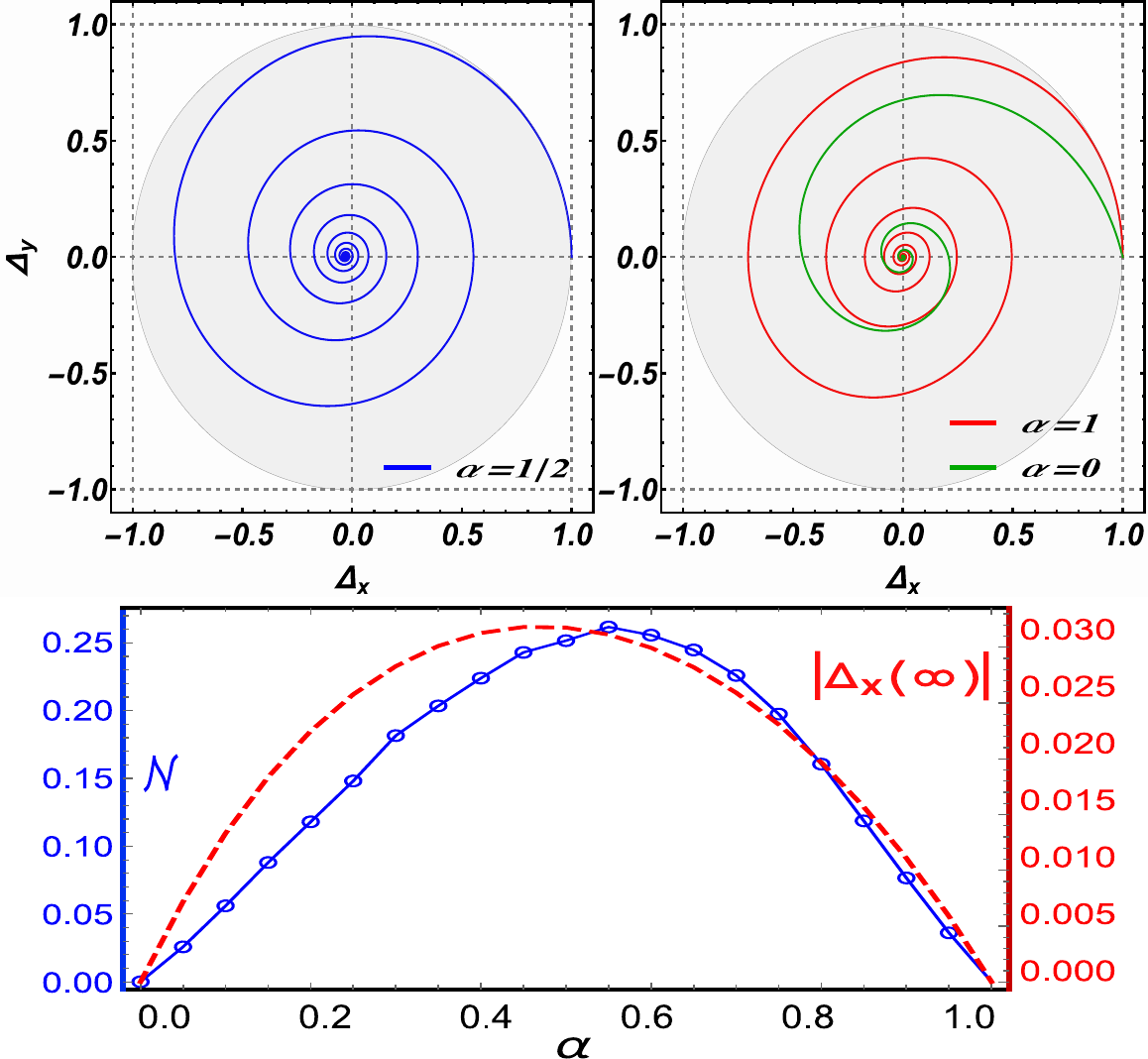}
		\caption{Top panel: Qubit coherence dynamics visualized as a path on the Bloch sphere equator,  where the left (right) panel stands for $ \alpha=1/2 $ ($ \alpha=0$ and $ \alpha=1$). Bottom panel: plots of Non-Markovianity (Blue line) and the absolute value of $ \Delta_x (t\to\infty) $ (red-dashed line) versus $\alpha$. Moreover, we set $ T=0.2~\omega_c $, $\varepsilon=0.5~\omega_c$, and $\eta=0.05$.}
		\label{fig: 1}
	\end{center}
\end{figure}
Figure (\ref{fig: 1}) illustrates the delicate interplay between the qubit’s coherence dynamics and the degree of non-Markovianity. The top panel displays how the coupling structure, controlled by the parameter $ \alpha $, determines the trajectory of the probe's coherence on the Bloch sphere. As anticipated, when $ \alpha = 1/2 $, the probe maintains quantum coherence for a longer duration and eventually stabilizes at a nonzero steady-state value. In contrast, for the case of $ \alpha = 0 $ (dephasing) and $ \alpha = 1 $ (dissipative), coherence decays faster to zero. This behavior stems from the interference between diagonal and off-diagonal couplings, encoded in the cross terms proportional to $ \alpha(1-\alpha) $ in Eqs.~(\ref{eq:dSxp}--\ref{eq:dSzp}). These terms mediate a continuous exchange of information between the population and coherence sectors, generating a feedback mechanism in which the environment both induces transitions and partially restores lost coherence. Consequently, the coexistence of the two coupling channels suppresses decoherence and gives rise to steady-state or trapped coherence, a \textit{signature} of non-Markovian memory effects~\cite{addis2014coherence,lin2022space,smirne2019improving}. The bottom panel further supports this interpretation: the non-Markovianity measure $ \mathcal{N} $ exhibits pronounced values that coincide with the re-coherence regions in the dynamics. The residual long-time coherence $ |\Delta_x(\infty)| $ thus quantifies the persistent memory retained by the probe, indicating that noncommutative coupling can effectively stabilize coherence against environmental noise by exploiting bath-induced backflow and interference effects.
	
\noindent\textit{Thermometric performance---} It has been shown that quantum sensing of the sample’s temperature is fundamentally bounded by means of the quantum Cramér–Rao bound (QCRB), which sets the minimum achievable uncertainty in the estimation of any unknown parameter, temperature in our case. The QCRB is given by~\cite{braunstein1994statistical,paris2009quantum}
	\begin{eqnarray}\label{QCRB}
		\delta T \geq \frac{1}{\sqrt{M \mathcal{F}_Q(T)}},
	\end{eqnarray}
	where $ \delta T $ denotes the standard deviation in the temperature estimate, $ M $ is the number of independent measurements and $ \mathcal{F}_Q(T) $ defines the quantum Fisher information (QFI) associated with the parameter $ T $. Therefore, maximizing the QFI plays a central role in optimizing the sensitivity of quantum thermometry. The QFI is defined as $
	\mathcal{F}_Q(T) = \mathrm{Tr} \{ \hat{L}_T^2 \, \rho_T \}$~\cite{paris2009quantum}, where the symmetric logarithmic derivative (SLD) operator $ \hat{L}_T $ satisfies $
	\partial_T \rho_T = \frac{1}{2} ( \hat{L}_T \rho_T + \rho_T \hat{L}_T ).$
	This quantity quantifies the maximum information that can be extracted about the temperature $ T $ from the quantum state $ \rho_T $.\par
	For the adopted model and by using the master equation given in Eq.~(\ref{NME}), one can compute the QFI with respect to the temperature $ T $ as
	$
	\mathcal{F}_Q = \left( \frac{\partial \vec{\mathbf{\Delta}}}{\partial T} \right)^{\!T} \cdot \mathbf{M}^{-1} \cdot \left( \frac{\partial \vec{\mathbf{\Delta}}}{\partial T} \right)
	$,
	where  $ \mathbf{M}^{-1} $ defines the inverse of the QFI metric tensor for a qubit, that is:
	$
	\mathbf{M}^{-1} = \mathbb{I}_3 + \frac{\vec{\mathbf{\Delta}} \vec{\mathbf{\Delta}}^T}{1 - \lvert \vec{\mathbf{\Delta}} \rvert^2}
	$, such that $\mathbb{I}_3$ is the $3 \times 3$ identity matrix.\par
	Note that when performing the numerical calculation of the QFI, one needs to evaluate the first-order derivatives of \(\partial_T \boldsymbol{\Delta}\). In this paper, the derivative of an arbitrary temperature-dependent function \(\boldsymbol{f}(T)\) is computed numerically using the following finite-difference method~\cite{hofmann2014scaling}:
	\begin{eqnarray}
	\footnotesize\frac{d\boldsymbol{f(T)}}{dT} \approx \frac{-\boldsymbol{f}(T + 2\delta) + 8\boldsymbol{f}(T + \delta) - 8\boldsymbol{f}(T - \delta) + \boldsymbol{f}(T - 2\delta)}{12\delta},\nonumber\\
	\end{eqnarray}
	where \(\delta = 10^{-7} \times T\), which provides a good accuracy.
	
	\section{Results \& discussion}
	\label{R_and_D}
	The central aim of this work is to investigate how the structure of the probe–sample interaction, characterized by the mixing parameter $ \alpha $~(Eq.~\ref{int-oper}), influences the performance of quantum thermometry in a nonequilibrium setting. Indeed, by varying $ \alpha $, we interpolate between purely dephasing thermometer~\cite{razavian2019quantum,albarelli2023invasiveness,yuan2023quantum} and purely dissipative thermometer~\cite{xu2023non,zhang2025low,ullah2025single,aiache2025quantum}, thereby exploring a broad class of hybrid interactions that combine coherence-preserving and energy-exchange effects. Our main goal is to evaluate the QFI associated with the probe's state. In what follows, we analyze the behavior of QFI as a function of $ \alpha $ at various fixed evolution times. This allows us to identify how nonequilibrium effects and the nature of the coupling jointly affect the sensitivity of the probe.\par
	\begin{figure}[H]
		\begin{center}
			\includegraphics[scale=.47]{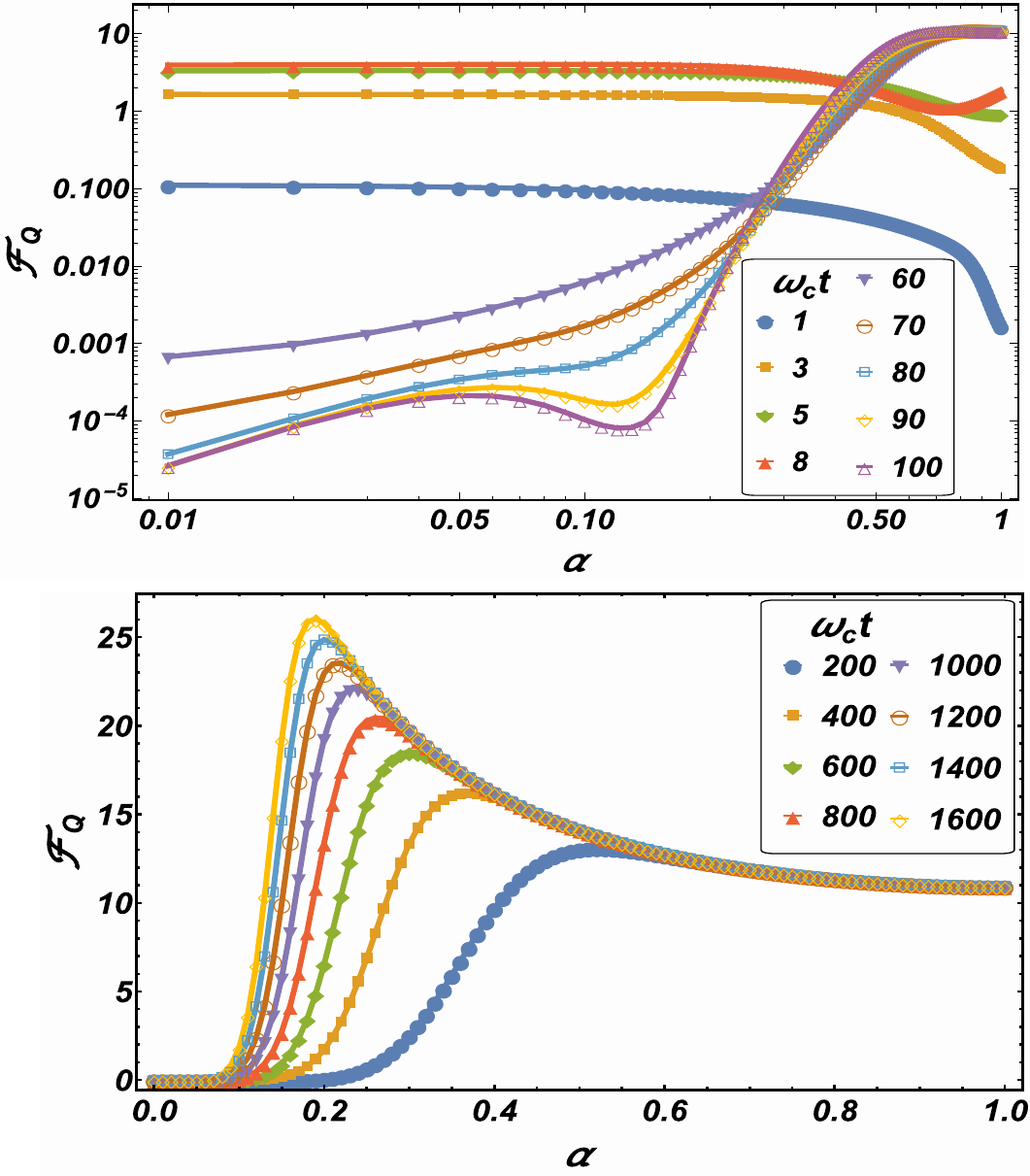}
			\caption{Top panel: QFI as a function of the mixing parameter $\alpha$ for fixed short and intermediate times.  Bottom panel: QFI at large times. All other parameters are the same as in Fig.~(\ref{fig: 1}).}
			\label{fig: 2}
		\end{center}
	\end{figure}
	Figure~(\ref{fig: 2}) exhibits the QFI as a function of the mixing parameter \( \alpha \) for various evolution times \( \omega_c t \). The top panel highlights the behavior at short to intermediate times. At a short interval of time, the temperature information is predominantly encoded in the nonequilibrium state of the probe via pure dephasing interactions (\( \alpha \to 0 \)). However, as \( \alpha \) increases toward unity, the sensitivity of the probe to temperature variations decreases, indicating that the off-diagonal interactions are less effective in this regime. However, at intermediate times, increasing \( \alpha \) leads to a noticeable improvement in sensitivity, up to a saturation point where the probe reaches thermal equilibrium with the environment, and no further information can be extracted.\par
	In contrast, the lower panel of Fig.~\ref{fig: 2} reveals that the QFI exhibits a clear enhancement at intermediate values of $ \alpha $, corresponding to noncommutative coupling configurations. This behavior indicates that the coexistence of dephasing and dissipative channels facilitates a more efficient extraction of temperature information from the bath. This enhancement originates from the interference between population and coherence dynamics, which allows the probe to simultaneously access both the energy exchange and phase fluctuation sectors of the bath. In this regime, the system operates far from thermal equilibrium, and the backflow of information characteristic of non-Markovian dynamics enables partial recovery of lost sensitivity during evolution. Such nonequilibrium feedback amplifies the temperature dependence of the populations and coherences, resulting in improved thermometric precision beyond the limits attainable in purely dephasing or purely dissipative interactions. This mechanism is reminiscent of recent findings in open quantum transport~\cite{duan2020unusual}, where noncommutative couplings and structured environments in a generalized spin-boson model are shown to significantly boost energy transport and thermal rectificationvia purely quantum effects. Therefore, the observed enhancement in QFI at intermediate $ \alpha $ reflects an interplay between coherence preservation and energy exchange, establishing noncommutative coupling as a resource for nonequilibrium quantum thermometry.\par
	
	In what follows, we focus on the early-time evolution of the probe and the scaling behavior at low temperatures. Measuring the temperature of a sample in the low-temperature regime is particularly challenging due to the vanishing of the QFI, which leads to a diverging quantum Cramér-Rao bound, i.e., $ \lim_{T \to 0} \mathcal{F}_Q(T) = 0 $. However, under the assumptions of weak coupling and the Born-Markov approximation, it has been shown that the QCRB diverges exponentially at low temperatures as~\cite{paris2015achieving,potts2019fundamental} 
	\begin{equation}
		\delta T^2 \geq \frac{2 T^4 e^{\varepsilon /T}}{M \varepsilon ^2}.
	\end{equation}
	The above expression clearly indicates that for $ T \to 0 $, the uncertainty in estimating the temperature increases rapidly. These limitations motivate the exploration of beyond-Markovian dynamics, where memory effects can alter the scaling behavior \cite{jorgensen2020tight}. In particular, we will examine how our non-Markovian master equation can enhance the QFI in the low-temperature regime.\par
	\begin{figure}[H]
		\begin{center}
			\includegraphics[scale=.44]{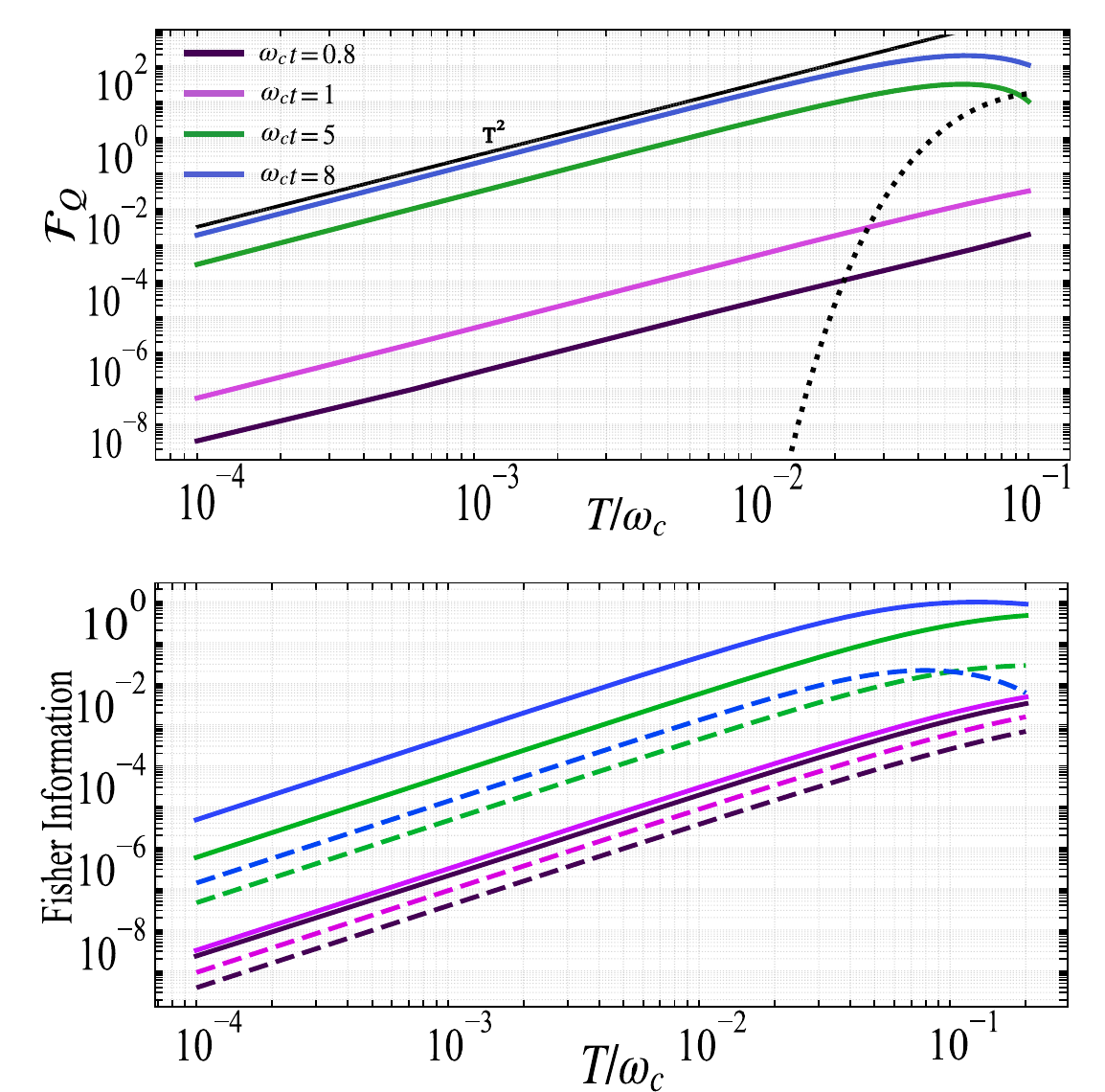}
			\caption{Top panel: QFI as a function of temperature for different probing times. The black solid line shows the $T^2$ scaling. The dotted line indicates the Fisher information obtained from the steady state of a secular Born–Markov master equation, which exhibits exponential scaling at low temperatures. Bottom panel: Fisher information obtained from coherence-based measurements (solid lines) and population-based measurements (dashed lines). Moreover, we set $\alpha = 1/2$, $\varepsilon = 0.5\omega_c$, and $\eta = 0.05$.}
			\label{fig: 3}
		\end{center}
	\end{figure}
	The top panel of Fig.~(\ref{fig: 3}) shows the non-equilibrium QFI versus the sample temperature for different probing times, in the presence of both diagonal and off-diagonal couplings ($ \alpha=1/2 $). We observe that non-Markovian dynamics can enhance the low-temperature scaling, where a non-vanishing QFI persists as $T \to 0$ and scales as $\mathcal{F}_Q \propto T^2$, in contrast to the Markovian case, where the QFI vanishes at low temperatures ($T \to 0$, black-dotted line). Interestingly enough, this favorable scaling behavior is achieved at early times~\cite{jorgensen2020tight} and under weak coupling, i.e., through \textit{nondestructive} probing.
	The $ \mathcal{F}_Q\propto T^2 $ low-temperature scaling has already been proven in a system of a fermionic tight-binding chain with access to only two sites~\cite{potts2019fundamental}, and for a quantum Brownian probe strongly coupled to an infinite bath of harmonic oscillators~\cite{correa2017enhancement}. It is worth mentioning that in this setup, the low-temperature scaling of the QFI also holds for a \textit{gapless} thermometer. In Ref.~\cite{hovhannisyan2018measuring}, it has been shown that for $\varepsilon = 0$ and $T/\omega_c \ll 1$, the QFI diverges as $\mathcal{F}_Q \propto 1/T^2$.\par
	Now, let us examine the Fisher information to identify the (\textit{sub})\textit{optimal} thermometer observable $\hat{\mathcal{O}}$ with the associated temperature uncertainty given in \cite{paris2009quantum,mehboudi2019using}
	\begin{equation}
		\delta T(\hat{\mathcal{O}}) \geq \frac{1}{\sqrt{M \mathcal{F}_C(\hat{\mathcal{O}})}},
	\end{equation}
	where $ \mathcal{F}_C(\hat{\mathcal{O}}) $ reflects the Fisher information associated with the performed measurement on the observable $ \hat{\mathcal{O}} $. The Fisher information serves as a precision quantifier for a specific measurement. Whereas the QFI determines the fundamental lower bound on the estimation error, regardless of the measurement performed. For a two-level system, the Fisher information associated with a measurement reads~\cite{mitchison2020situ}
	\begin{equation}
		\mathcal{F}_C(\hat{\mathcal{O}})=\dfrac{1}{\braket{\Delta \hat{\mathcal{O}}^2}}\bigg(\dfrac{\partial\braket{\hat{\mathcal{O}}}}{\partial T}\bigg)^2,
	\end{equation}
	where $ \braket{\hat{\mathcal{O}}} $ and $ \braket{\Delta \hat{\mathcal{O}}^2} $ represent the mean and variance of the measured observable $ \hat{\mathcal{O}} $. In this work, we will focus on comparing measurements of coherences and populations-difference, i.e., measuring $ \sx $ and $ \sz $, receptively. Thus, Fisher information reads $ \mathcal{F}_C(\sx)=(\partial_T \Delta_x)^2/(1 - \Delta_x^2) $ and for populations-difference ($ \sz $), we simple replace $\Delta_x$ with $ \Delta_z $.\par
	
	%In Fig.~(\ref{fig: 3})~(bottom), we show the Fisher information from coherence-based (solid) and population-based (dashed) measurements in the low-temperature and early-time regimes. Both observables exhibit a $T^2$ scaling as $T \to 0$, consistent with the QFI behavior. In this regime, coherence measurements yield higher Fisher information values, making it a better, though still \textit{suboptimal}---choice for extracting temperature information in practical settings.\par
	In Fig.~\ref{fig: 3}(bottom), we compare the Fisher information obtained from coherence-based (solid) and population-based (dashed) measurements in the low-temperature and short-time regimes. Both observables exhibit a quadratic scaling as $ T \to 0 $, consistent with the behavior of the QFI. The superiority of coherence-based measurements at early times arises because quantum coherences encode the immediate response of the probe to temperature-induced phase fluctuations, before dissipative energy exchange with the bath becomes dominant. In this transient regime, off-diagonal elements of the density matrix carry direct information about the noise spectrum and its temperature dependence, allowing a fast and sensitive estimation process. In contrast, population-based observables respond on longer relaxation timescales, reflecting slower thermalization dynamics.\\
	Moreover, short-time non-equilibrium thermometry can be both highly informative and minimally invasive~\cite{albarelli2023invasiveness}, since measurements performed before full equilibration disturb the system less while still capturing the essential thermal signatures. In this sense, the enhanced sensitivity of coherence-based measurements observed here not only reflects their access to fast non-Markovian correlations but also aligns with the principle of optimal, non-invasive quantum thermometry, where precision and minimal disturbance coexist in the early-time regime. This highlights the operational relevance of coherence measurements and provides a practical route to experimentally feasible, nonequilibrium quantum thermometers.\par
	We note that at sufficiently low temperatures, the (quantum) Fisher information deviates from the Markovian result, even for weak coupling. This highlights that faithfully extracting temperature in this regime requires accounting for system–environment correlations, which are absent in simple dissipation models such as those described by the GKSL master equation~\cite{lindblad1976generators,gorini1976completely}.
	
	\section{Experimental Outlook}
	\label{Experimental_Outlook}
	%The present work introduces a tunable system–bath coupling operator $ \hat{\sigma}_{\alpha} = (1 - \alpha)\, \sz + \alpha\, \sx $ that continuously interpolates between pure dephasing and pure dissipation, thereby unifying the two major coupling paradigms within quantum thermometry. This unified treatment reveals qualitatively new physical insights: in particular, we demonstrate that intermediate values of $ \alpha $ enable enhanced temperature sensitivity, manifested via non-Markovian memory effects and a low-temperature scaling $ \mathcal{F}_{T} \sim T^{2} $ that sharply contrasts with the exponential suppression predicted under standard Born–Markov approaches. By revealing optimal coupling structures for thermometric precision, our work goes beyond formal extension of prior models and uncovers controllable resources (non-commutativity of couplings + non-Markovianity) for quantum sensing.	
	From an experimental perspective, the coupling structure embodied in $ \hat{\sigma}_{\alpha} $ is not purely theoretical. For example, in superconducting-circuit architectures an inductively shunted transmon qubit has been demonstrated with flux-tunable switching between transverse ($ \sx $) and longitudinal ($ \sz $) coupling to a harmonic mode~\cite{richer2017inductively}. Moreover, hybrid spin-photon systems have achieved parametric longitudinal coupling ($ \sz $) between a semiconductor spin qubit and a high-impedance resonator, underscoring the feasibility of engineered qubit–bath interactions beyond standard transverse ($ \sx $) coupling~\cite{bottcher2022parametric}. 
	%In such platforms one may implement the parameter $ \alpha $ by tuning drive amplitude, flux bias, or coupling axis, and then perform state tomography of the qubit to extract populations and coherences. By scanning temperature in the cryogenic regime and comparing temperature estimation error under different coupling angles $ \alpha $, one could directly test our predicted enhancement of Fisher information and the non-Markovian advantage. Thus our results provide a realistic framework for next-generation quantum thermometry experiments in solid-state platforms with engineered non-commutative noise channels.
	In such platforms, the coupling strength and orientation can be tuned through drive amplitude, flux bias, or coupling-axis control, and qubit state tomography can be performed~\cite{aasen2024readout} to extract both populations and coherences. By sweeping the sample temperature in the cryogenic regime~\cite{lvov2025thermometry} and comparing the achieved temperature estimation error for different coupling configurations, one could directly test the Fisher-information enhancement and the influence of non-Markovian dynamics.
	
	\section{Conclusions}
	\label{Concl}
	The present work introduces a tunable system–bath coupling operator
	$ \hat{\sigma}_{\alpha} = (1 - \alpha) \hat{\sigma}_z + \alpha \hat{\sigma}_x $
	that continuously interpolates between pure dephasing and pure dissipation, thereby unifying the two major coupling paradigms within quantum thermometry. This unified treatment reveals qualitatively new physical insights: in particular, we demonstrate that intermediate values of $ \alpha $ enable enhanced temperature sensitivity, manifested via non-Markovian memory effects and a low-temperature scaling of the QFI that sharply contrasts with the exponential suppression predicted under standard Born–Markov approaches. By revealing optimal coupling structures for thermometric precision, our work goes beyond formal extension of prior models and uncovers controllable resources (non-commutativity of couplings + non-Markovianity) for quantum sensing.
	
	Moreover, we analyzed the scaling behavior of the QFI at low temperatures and short times for the co-existing diagonal and off-diagonal couplings. In this regime, we concluded that $\mathcal{F}_Q\propto T^2$, which is consistent with fundamental thermodynamic limits~\cite{potts2019fundamental,correa2017enhancement,jorgensen2020tight}. At sufficiently low temperatures, deviations from the Markovian prediction appeared even for the weak coupling regime. Further, we showed that in this regime a measurement based on the coherence~\cite{ullah2023low,aiache2024harnessing} of the thermometer is \textit{suboptimal} to extract thermal information from the sample.\par
	Beyond the theoretical implications of our model, the present framework provides a physically grounded route toward experimentally relevant quantum thermometry. The generalized noncommutative coupling operator can be realized in controllable settings where both diagonal and off-diagonal noise channels are routinely engineered. In such environments, the interference between dephasing and dissipative processes identified in this work could be tuned through external controls~\cite{richer2017inductively}
	%or engineered reservoirs, 
	enabling the observation of coherence trapping and enhanced thermometric sensitivity in nonequilibrium regimes. These findings indicate that noncommutative system–bath couplings represent a versatile resource for improving temperature estimation in realistic quantum devices operating at cryogenic or mesoscopic scales~\cite{lvov2025thermometry}. 
	%This opens a promising avenue for experimentally benchmarking nonequilibrium quantum thermometry beyond the Markovian domain.

	\bibliographystyle{apsrev4-1}   % or any style you want
	\bibliography{references}  % references.bib

\end{document}